\documentclass[twoside]{article}
\usepackage{fleqn,espcrc2}
\usepackage{graphicx}
\usepackage{here}

\newcommand{\AmS}{{\protect\the\textfont2
    A\kern-.1667em\lower.5ex\hbox{M}\kern-.125emS}}										
\def\beq{\begin{equation}}
\def\eeq{\end{equation}}
\def\bea{\begin{eqnarray}}
\def\eea{\end{eqnarray}}
\def\bq{\begin{quote}}
\def\eq{\end{quote}}

\def\nnb{\nonumber}
\def\ga{\left(}
\def\dr{\right)}
\def\aga{\left\{}
\def\adr{\right\}}

\def\rar{\rightarrow}
\def\lrar{\Longrightarrow}

\def\lrar{\leftrightarrow}
\def\nnb{\nonumber}
\def\la{\langle}
\def\ra{\rangle}
\def\nin{\noindent}
\def\ba{\begin{array}}
\def\ea{\end{array}}

\def\b{$\bullet~$}
\def\als{\alpha_s}

\def\gg2{ \la\alpha_s G^2 \ra}
\def\gg3{g^3f_{abc}\la G^aG^bG^c \ra}
\def\ggg4{\la\als^2G^4\ra}

\def\one{{\rm 1\kern -.9mm l}}                             %

\def\beq{\begin{equation}}
\def\eeq{\end{equation}}
\def\beqa{\begin{eqnarray}}
\def\eeqa{\end{eqnarray}}
\newcommand{\ena}{\end{eqnarray}}

\DeclareMathAlphabet{\eusm}{U}{}{}{}
\SetMathAlphabet\eusm{normal}{U}{eus}{m}{n}
\SetMathAlphabet\eusm{bold}{U}{eus}{b}{n}
\DeclareMathAlphabet{\mathpzc}{OT1}{pzc}{m}{it}

\title
{\bf{\boldmath
{\Large A Fresh Look into the Neutron EDM and Magnetic Susceptibility} }}

\author{Stephan Narison\thanks{Email: snarison@yahoo.fr}  \address {\footnotesize Laboratoire
de Physique Th\'eorique et Astroparticules, Universit\'e
de Montpellier II, Case 070, Place Eug\`ene
Bataillon, 34095 - Montpellier Cedex 05,
France},
}

\begin{document}

\pagestyle{myheadings}
\markright{ }
\begin{abstract}
\noindent
We reexamine the estimate of the neutron Electric Dipole Moment (NEDM)  from chiral and QCD spectral sum rules (QSSR) approaches. In the former, we evaluate the pion mass corrections which are about 5\% of the leading Log. results. However, the chiral estimate can be affected by the unknown value of the renormalizaton scale $\nu$. For QSSR, we analyze the effect of the nucleon interpolating currents on the existing predictions. We conclude that previous QSSR results are not obtained within the optimal choice of these operators, which lead to an  overestimate of these results by about a factor 4.  The weakest upper bound $|\theta|\leq 2\times 10^{-9}$ for the strong $CP$-violating angle is obtained from QSSR,
while the strongest upper bound $|\theta|\leq 1.3\times 10^{-10}$ comes from the chiral approach evaluated at the scale $\nu=M_N$. 
We also re-estimate the proton magnetic susceptibility, which is an important input in the QSSR estimate of the NEDM. 
\end{abstract}
\maketitle

\vspace{1cm}
\section{Introduction} \label{intro}
The Lagrangian of Yang-Mills theory contains, in addition to the usual
term, also a topological term:
\begin{equation}
{\cal L} = - \frac{1}{4} F^{a}_{\mu \nu} F^{a \mu \nu} - \theta q (x)~,
\label{topo}
\end{equation}
where $ q(x)$ is the topological charge density given by:
\begin{equation}
q(x) = \frac{g^2}{32 \pi^2} F_{\mu \nu}^{a} {\tilde{F}}^{a \mu \nu}
~~~:~~~ {\tilde{F}}^{\mu \nu} = \frac{1}{2} \epsilon^{\mu \nu \rho
  \sigma} F_{\rho \sigma}~.
\label{q}
\end{equation}
The additional term violates the invariance under $CP$. This is called
strong $CP$ violation \cite{PECCEI}  to distinguish it from the $CP$ violation present in the
weak and electromagnetic sector of the Standard Model. 
Experiments, however, do not show any violation of strong $CP$ and require a
very small value for $|\theta| < 2\times 10^{-9}$ as we shall see later on.
In this paper we shall determine the dependence of physical
quantities on $\theta$ and study the processes that violate  strong
$CP$. For this purpose, we reevaluate the neutron electric dipole moment
(NEDM) which depends on the $\pi NN$ coupling which violates CP.
\section{Improved chiral estimate of the NEMD }
An elegant way of doing this is to use
the low energy effective Lagrangian of QCD that contains the fields of the 
pseudoscalar mesons and baryons instead of the original quarks and
gluons. This is due to the fact that in the effective Lagrangian the
effect of the axial $U(1)$ anomaly is explicitly displayed and because
of this the amplitudes for the hadronic processes can be easily computed.
This Lagrangian cannot be
explicitly derived from the fundamental QCD Lagrangian as in the 
$CP^{N-1}$ model (see e.g. Ref.~\cite{DV2,divecchia} and  
References therein), but can only be
constructed  requiring that it has the same anomalous and
non-anomalous symmetries of the fundamental QCD Lagrangian. The expressions
of the QCD effective  Lagrangian describing pseudoscalar mesons and including the $U(1)$ anomaly are given in  \cite{W1,GV,DV1,CVW,RST,DGV,W2,BALUNI} and  the review in
Ref.~\cite{DV2,divecchia}.
\subsection*{\b Estimate of the $\pi NN$ couplings}
The $\pi NN$ couplings are defined as \footnote{We follow the normalizations of \cite{CVW} but adding an overall $\sqrt{2}$ factor for charged pion fields.}:
\beq
{\cal L}_{\pi NN}=\sqrt{2}\pi\bar N{\bf\tau}\ga i\gamma_5 g_{\pi NN}+\bar g_{\pi NN}\dr N~,
\eeq
where ${\bf\tau}$ are isospin Pauli matrices.

The $CP$-conserving coupling is well measured :
\beq
g_{\pi NN}=13.4.
\label{eq:CPcons}
\eeq

The $CP$-violating coupling $\bar g_{\pi NN}$ can be obtained using an effective Lagrangian
approach, though this approach 
for estimating the coupling can be questionable. 
It reads \cite{CVW,divecchia}:
\bea {\bar{g}}_{\pi NN} = - \frac{m_u m_d \theta}{ f_{\pi}(m_u +m_d ) (m_s
 -m)} \left(  m_{\Xi}
-m_{\Sigma} \right)\times \nnb\\
\left[ 1 + \frac{3m (m_{\Sigma} - m_{\Lambda}) }{2 (m_s
 -m) m_N} \right]
\label{gbarb}
\eea
with: $2m\equiv (m_u+m_d)$.
We shall use the ChPT mass ratio \cite{LEUT}:
\beq
r_3\equiv  \frac{2m_s}{ (m_u+m_d)}= 24.4\pm 1.5~,
\eeq
which is confirmed by the QSSR estimates of the quark mass absolute values in units of MeV \cite{SNQ,SNB}:
\beq
m_u(2)=(2.8\pm0.2)~,~~~m_d(2)=(5.1\pm0.4)~,
\eeq
and:
\beq
~~~ m_s(2)=(96.3\pm17.5)~.
\eeq
Using $f_\pi= 92.46$ MeV, one can deduce: 
\beq
\bar g_{\pi NN}= (0.0282\pm 0.0071) \theta~,
\label{eq:CPviol}
\eeq
which is relatively small compared to $g_{\pi NN}$. One can also note that the chiral correction is about 0.5\% of the leading order expression given in Ref.~\cite{CVW}. We have added an error of about 25\% 
as a guess of the systematics of the method based on its deviation for predicting the value of the measured $CP$-conserving coupling \footnote{See however Ref. \cite{OHTA}.}.
\subsection*{\b The NEDM from chiral and ChPT approaches}
\begin{center}
\begin{figure}[here]
\includegraphics[width=.4\textwidth]{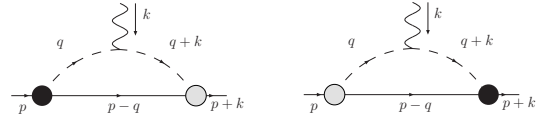}
\caption{1st class of diagrams contributing to $D_N$.}
\label{Fig1stef}
\end{figure}
\end{center}

\begin{center}
\begin{figure}[here]
\includegraphics[width=.4\textwidth]{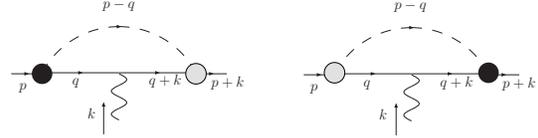}\caption{2nd class of diagrams contributing to $D_N$.}
\label{Fig2stef}
\end{figure}
\end{center}

To first order in $\theta$, the neutron electric dipole moment (NDEM) $D_N$ is given by:
\bea
V_\mu&\equiv& {\cal T}\la n(p_f)|J_\mu(0)~i\int d^4x\delta {\cal L}_{CP}(x)|n(p_i)\ra\nnb\\
&=& -\ga iD_N\dr\bar u(p_f) \sigma_{\mu\nu} k^\nu \gamma_5 u(p_i)+{\cal O} (k^2)~,
\eea
where : $k\equiv p_f-p_i$ is the photon momentum and:
\beq \delta {\cal L}_{CP}(x)=\bar q(i\gamma_5){\cal A}q~,
\eeq
with {\cal A} a 3-dimension hermitian matrix acting on the flavour space (u,~d,~s) and 
$
\sigma_{\mu\nu}\equiv  ({1/ 2}) \ga \gamma_\mu\gamma_\nu -\gamma_\nu\gamma_\mu \dr~.
$
In order to extract $D_N$, we use the Gordon decomposition for the axial current:
\beq
\bar u(p+k)\gamma_5\sigma_{\mu\nu}k^{\nu} u(p)=...+
\bar u(p+k)\gamma_5(2p+k)_\mu u(p).
\eeq
The lowest order contribution to $D_N$ comes from the diagrams in Figs \ref{Fig1stef} and \ref{Fig2stef}. Using 
the expression of the $\pi\pi\gamma$ vertex :
\beq
\la \pi^-(p+k)|J_\mu|\pi^-(p)\ra=-(2p+k)_\mu +{\cal O}(k^2)~,
\eeq
the evaluation of the previous diagrams can be expressed as:
\bea\label{eq:edmchiral}
D_N=&&(-1)^3\bar g_{\pi NN}g_{\pi NN}\ga {2M_N\over M_N^2}\dr \ga{1\over 16\pi^2}\dr \times \nnb\\
&& 2\aga I^{(1)}+I^{(2)}\adr
\eea
where $ I^{(n)}$ are integrals over Feynman parameters coming from Figs \ref{Fig1stef} and \ref{Fig2stef}.
Fig \ref{Fig1stef}  gives:
\bea\label{eq:i1integral}
I^{(1)}&\equiv& \int_0^1 xdx \int_0^1 dy~ {x(1-y)\over x^2(1-y)^2+a\Big{[} 1-x(1-y)\Big{]}}
\nnb\\
&=& -1-{\log {a}\over 2}\ga 1-a\dr +f(a)
\eea
with:
\bea
 f(a)&=&
{a\ga 3-a\dr \over \sqrt{a(4-a)}}
 \Big{\{}\arctan{\Big{[} {2-a\over \sqrt{a(4-a)}}\Big{]}}+\nnb\\
&& \arctan{\Big{[} {a\over \sqrt{a(4-a)}}\Big{]}}\Big{\}}\nnb\\
 &\simeq& {3\over 4}\ga\pi\sqrt{a}-a\dr -{5\over 32}\pi a^{3/2}+\ldots
~,
\eea
and where: 
\beq
a\equiv  \ga m_\pi/ M_N\dr^2~.
\eeq 
One can notice that the term $\log{a}$ appears naturally in the unexpanded full expression without
an arbitrary choice of the cut-off $M_N$. 
One can also note the non-analytic terms $\sqrt{a}$ and $a^{3/2}$.
In the same way, Fig \ref{Fig2stef} gives:
\bea
I^{(2)}&\equiv& \int_0^1 xdx\int_0^1 dy  {x\over x^2+a(1-x)}\nnb\\
&=& 1+{(-2+a)a\over \sqrt{(4-a)a}}\Big{\{} \arctan{(2-a)\over  \sqrt{(4-a)a}}-\nnb\\
&& \arctan{a\over  \sqrt{(4-a)a}}\Big{\}} -{a\over 2}\log {a}\nnb\\
&=& 1-{\pi\over 2}\sqrt{a}+{a\over 2}\ga 3-\log{a}\dr+{3\over 16}\pi a^{3/2}.
\label{eq:i2integral}
\eea

 \subsection*{\b Results and discussions}
 Adding the previous  Feynman integrals $I^{(1)}$ and $I^{(2)}$ [ Eqs. (\ref{eq:i1integral}) and (\ref{eq:i2integral})] into the expression of NEDM in Eq. (\ref {eq:edmchiral}), the NEDM from chiral approach reads:
 \bea
 D_N\vert_{\rm chiral}&=&{\bar g_{\pi NN}g_{\pi NN} \over M_N} \ga{1\over 4\pi^2}\dr \times \Bigg{\{}
  -{\log {a}\over 2}+\nnb\\
 && 0+{1\over 4}\ga\pi \sqrt{a}+3a\dr +\ldots  \Bigg{\}}~.
 \eea
The leading-log term agrees with the original result in Ref~ \cite{CVW}. The cancellation of the constant terms have been also noticed in \cite{Pich}. In addtion, we also have a cancellation of the $a\log{a}$ term, which implies small mass corrections (about 5\%). Our result demonstrates the accuracy of the leading-log approximation used in Ref.~\cite{CVW}. It also shows that the scale, appearing in the leading-$\log{a}$ term, is the mass of the nucleon but not any arbitrary cut-off  scale, because this term appears  before the expansion in $a$
[see Eq. (\ref{eq:i1integral})]. 
This leads to the prediction:
\bea
D_N\vert_{\rm chiral}&=& (20\pm 5)\times 10^{-3}\theta ~{\rm GeV}^{-1}~,\nnb\\
&=& (40\pm 10)\times 10^{-17}\theta~{\rm cm}~.
\label{eq:dnchiral}
\eea
However, within Chiral Perturbation Theory (ChPT), the addition of counterterms in the effective Lagrangian induces a $\log (M_N/\nu)^2$-term and leads to the (renormalized) NEDM expression \cite{Pich}:
 \bea
 D_N(\nu)\vert_{\rm ChPT}={\bar g_{\pi NN}g_{\pi NN} \over M_N} \ga{1\over 4\pi^2}\dr 
 \Bigg{\{} \log{\nu\over m_\pi }+k\Bigg{\}}
 \eea
 where $\nu$ is an arbitrary hadronic scale and $k$ an unknown constant.  This result indicates that,
 only the coefficient of the $\log m_\pi$  term is model independent. 
 
 For a conservative estimate and
 for a model independent result, we keep only the $\log m_\pi$  term and move, like in \cite{Pich},  the scale $\nu$ above $m_\pi$ from the value of the constituent quark mass, which we take to be about $(M_N/3)$ (within a 30\% error), to  $M_N=940$ MeV. Using the values of the parameters in Eqs. (\ref{eq:CPcons}) and (\ref{eq:CPviol}), 
 one obtains, in units of the electric charge $e$:
\bea
D_N(\nu)&=& (2.7\sim 23.8)\times 10^{-3}\theta ~{\rm GeV}^{-1}~,\nnb\\
&=& (5.4\sim 47.6)\times 10^{-17}\theta~{\rm cm}~,
\label{eq:dnChPT}
\eea
where the range takes into account the assumed 25\% systematic uncertainties for estimating $\bar g_{\pi NN}$ and the assumed 30\% one for the value of the quark constituent mass. The value in Eq. (\ref{eq:dnchiral}) corresponds to the value $\nu=M_N$ and where the small chiral mass corrections have been included. One can notice that the width of the range depends strongly on the unknown value of $\nu$, and where a fine tuning is obtained for its low values. 
\section{NEDM and Magnetic Susceptibility from QSSR }
Alternative estimates of the NEDM have been done using QCD spectral sum rules \cite{ritz}. We shall re-examine these results in this section and present some alternative new sum rules.  \\
The analysis is based on the baryon two-point correlator put in a background with nonzero $\theta$ and electromagnetic field $F_{\mu\nu}$:
\bea
S(q^2) \vert_{\theta,~F}=i\int d^4x~e^{iqx}\la 0\vert {\cal T} N(x)\bar N(0)\vert 0 \ra \vert_{\theta,~F}
\eea\label{eq:2point}
$N$ is the nucleon operators which can be written in general as~\footnote{In order to help the reader, 
we use the same notations and normalizations as in Ref. \cite{ritz}, which is the same as the ones in \cite{ioffe}. The normalization of the QCD correlator differs by a factor 8 from \cite{dosch,dosch1}. } :
\beq
N(x)\equiv 2\aga\ga \psi C\gamma^5\psi\dr\psi + b \ga \psi C\psi\dr\gamma^5\adr~,
\eeq
where $C$ is the charge conjugate; $\psi$ is  the quark field; $b$ is (a priori) an arbitrary mixing between the two operators: $b=0$ in the non-relativistic limit,
which is the choice used in different lattice calculations \cite{LEIN}. Originally, Ioffe \cite{ioffe} has used the choice $b=-1$ in the 1st QSSR applications to the nucleon channel, which he has justified for a better convergence of the QCD series in the OPE. In \cite{dosch,dosch1}~\footnote{For reviews, see e.g. \cite{SNB,LEIN}.}, a more general analysis has been performed by letting $b$ as a free parameter and then looks for the $b$-value where the result is less sensitive to the variation of $b$. From the overall fit using different form of the sum rules, the optimal result for the nucleon mass and residue have been obtained for $b=-1/5$, which can be qualitatively understood by taking the zero of the derivative in $b$ when retaining only the lowest order contribution. \\
For the analysis of the NEDM, Ref. \cite{ritz} has used a value $b=1$, which differs completely from the previous choices. Within this choice, the authors impose the zero coefficient of an non-analytic mass- Log. in the 
next $1/Q^2$-corrections $(q^2\equiv -Q^2)$ to the lowest order contribution. However, the vanishing of the mass- Log. to leading order does not (a priori) guarantee the absence of this contribution to higher orders. In the following, we shall test the stability of the existing results versus the variation of $b$. 
\subsection*{\b Expression of the two-point correlator }
For the forthcoming analysis, we shall keep the coefficient of the term
$
\{ \tilde F_{\mu\nu}\sigma^{\mu\nu}, \gamma_{\mu} q^\mu\}~,
$
in the Lorentz decomposition of the nucleon two-point correlator given in Eq. (\ref{eq:2point}) (some alternative choices are also possible). The QCD expression reads \cite{ritz} :
\bea
&S(Q^2) \vert_{\theta,~F}^{\rm th} = -\theta m^*\la \psi\psi\ra \ga {1\over 32\pi^2}\dr \Bigg{\{} -\chi C_0 \ln {Q^2\over \nu^2}+\nnb\\
&\Big{[}C_{2a}\ga \ln{Q^2\over \nu_\chi^2}-1\dr +C_{2b} \Big{]}{1\over Q^2}+{\cal O}\ga{1\over Q^4}\dr\Bigg{\}}
\eea
$\nu$ and $\nu_{\chi}$ correspond to an UV and a small mass arbitrary subtractions; $\la \bar\psi\psi\ra$ is the quark condensate. The coefficient-functions
are:
\bea
C_0= (b+1)^2(4e_d-e_u)~,\nnb\\
C_{2a}=-4(b-1)^2e_d\ga 1+{1\over 4}(2\kappa+\xi)\dr~,\nnb\\
C_{2b}=-{\xi\over 2}\Big{[}(4b^2-4b+2)e_d+(3b^2+2b+1)e_u\Big{]}~,
\eea\label{eq:qcd}
where $e_u,~e_d$ are the electric charge of the $u$ and $d$ quarks in units of e. $\chi,~\zeta,~\xi$ are the magnetic susceptibilities of the QCD condensates which encode the electromagnetic field dependence of the two-point correlator. In units of $e$, they are defined as \cite{IOFFE2}:
\bea
\la 0\vert \bar\psi \sigma_{\mu\nu} \psi \vert 0\ra \vert_{F}&=&e_q\chi F_{\mu\nu}\la  \bar\psi\psi\ra~,\nnb\\
g\la 0\vert \bar\psi {\lambda_a\over 2}G^a_{\mu\nu} \psi \vert 0\ra \vert_{F}&=&e_q\kappa F_{\mu\nu}\la  \bar\psi\psi\ra~,\nnb\\
g\la 0\vert \bar\psi \gamma_5{\lambda_a\over 2}G^a_{\mu\nu} \psi \vert 0\ra \vert_{F}&=&ie_q\xi F_{\mu\nu}\la  \bar\psi\psi\ra~,
\label{eq:magnetic}
\eea
where $g$ is the QCD coupling and $G^a_{\mu\nu}$ the gluon field strength. The size of these magnetic susceptibilities have been 
estimated in the literature using different methods. The values \cite{WYLER} :
\beq
\kappa\simeq -(0.34\pm 0.10)~,~~~~~~~~\xi\simeq -(0.74\pm 0.2)~,
\label{eq:xikappa}
\eeq
induce (a posteriori) small numerical corrections in the present analysis and will not be reconsidered. On the contrary, the dominant contribution comes from $\chi$, which is not known with a good accuracy:
\bea
\chi[{\rm GeV}^{-2}]= &&-8=-{N_c \over 4\pi^2f_\pi^2}:~{\rm Triangle~ anomaly } \cite{VAIN}\nnb\\
&&-8,~-6 ~~{\rm Laplace ~SR (LSR)}  \cite{IOFFE2,WYLER}\nnb\\
&&-3.3~~{\rm Light ~cone~ SR} \cite{BALL}~,
\eea
which we shall reconsider later on. Our analysis gives the value in Eq. (\ref{eq:chisr}), which is
in better agreement with the one obtained in \cite{IOFFE2,VAIN}.

The phenomenological part of the correlator can be parametrized in the zero width approximation by:
\bea
 S(q^2) \vert_{\theta,~F}^{exp}&=& {\lambda_N^2M_ND_N\over (q^2-M^2_N)^2}+{A\over q^2-M_N^2}+\nnb\\
 && ``{\rm QCD~ continuum}"~,
 \eea 
 where $\lambda_N$ is the nucleon coupling to its corresponding current; A is an arbitrary coupling which 
 parametrizes the single pole contributions; ``QCD continuum" stands from the QCD smearing of  excited state contributions and comes from the discontinuity of the QCD expression. 
\subsection*{\b Estimate of the  Magnetic susceptibility}
\begin{center}
\begin{figure}[here]
\includegraphics[width=.22\textwidth]{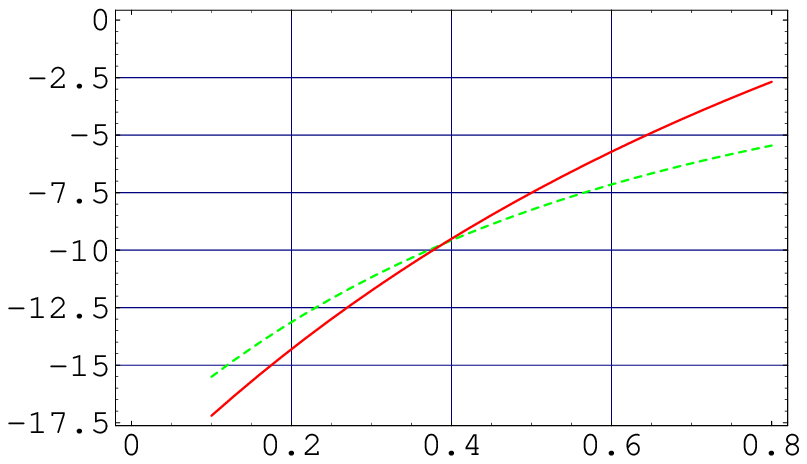}
\includegraphics[width=.22\textwidth]{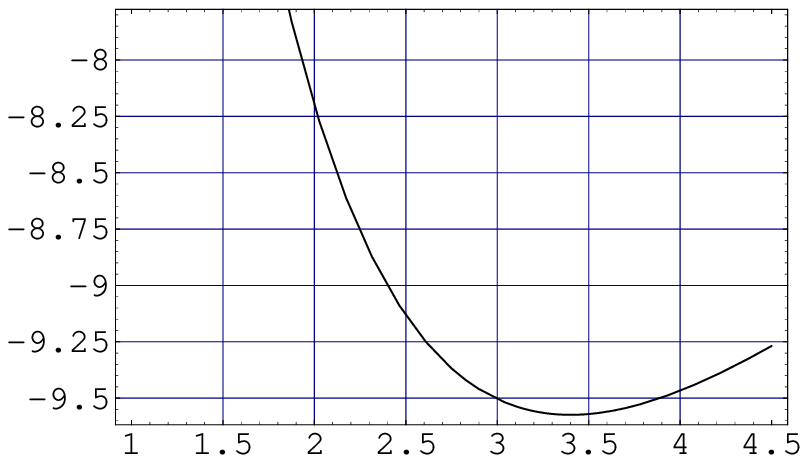}
\caption{Analysis of $\chi$ using LSR for b=-1: a) $\tau$-dependence for $t_c=3.$ GeV$^2$; b) $t_c$-dependence of the common solution in (a) for $\tau=.4$ GeV$^{-2}$.  }
\label{fig:chioffe}
\end{figure}
\end{center}

\nin
Considering the nucleon two-point correlator in Eq. (\ref{eq:2point}) in presence of an external electromagnetic field,
one can derive the following LSR (neglecting anomalous dimensions) for each invariants related to the structure
$(\sigma_{\mu\nu}\hat {p} + \hat{p}\sigma_{\mu\nu})$ and $i(p_\mu\gamma_\nu-p_\nu\gamma_\mu)$ \cite{IOFFE2} (hereafter referred as IS):
\bea\label{eq: sigma}
{\cal L}^\sigma_p&\equiv& e_uM^4(1-\rho_1)+{a^2\over 3M^2}\Bigg{\{}-(e_d+{2\over 3}e_u)+
\nnb\\
&&{1\over 3}e_u(\kappa-2\zeta)
 -2e_u\chi(M^2-{1\over 8}M_0^2)\Bigg{\}}\nnb\\
 &=&{1\over 4}\tilde\lambda_N^2 e^{-M_N^2/M^2}\ga {\mu_p\over M^2}+A_p\dr~,
\eea
\bea\label{eq: gamma}
{\cal L}^\gamma_p&\equiv& aM_N\Bigg{\{} e_u+{1\over 2}e_d+\nnb\\
&&{1\over 3}e_d\chi M^2\Big{[}(1-\rho_0)+{b\over 24M^4}\Big{]}\Bigg{\}}\nnb\\
&=& {1\over 4}\tilde\lambda_N^2 e^{-M_N^2/M^2}\ga {\mu^a_p\over M^2}+B_p\dr~,
\eea
where $M^2\equiv 1/\tau$ is the SR variable; $M_N=0.946$ GeV is the proton mass;  $a\equiv 4\pi^2|\la 0|\bar \psi\psi|0\ra|$; $b=4\pi\la 0|\alpha_s G^2|0\ra\simeq 0.87$ GeV$^{4}$ \cite{SNGLUE,SNB} are the quark and gluon condensates;  $M^2_0=0.8$ GeV$^2$ \cite{ioffe,dosch,dosch1,SNMIX,SNB} parametrizes the mixed quark-gluon condensate: $\la 0|\bar \psi \sigma_{\mu\nu}(\lambda^a/2) G_a^{\mu\nu} \psi|0\ra= M^2_0\la 0|\bar \psi\psi|0\ra$;  The QCD continuum contribution from a threshold $t_c$ is quantified as:
 \beq
 \rho_n= e^{-t_c\tau}\ga 1+t_c\tau+{(t_c\tau)^2\over 2}+...+{(t_c\tau)^n\over n!}\dr~;
 \label{eq:continuum}
 \eeq
 $A_p,~B_p$ are the single proton pole coupling to the two-point correlator, while $\tilde\lambda_N^2\equiv 32\pi^4\lambda^2_N$ is the coupling of the double proton pole; 
$\kappa$, $\zeta$ and $\chi$ have been defined in Eq. (\ref{eq:magnetic}); $\mu_p$ and $\mu_p^a$ are the proton magnetic and anomalous magnetic moments. 
The sum rules for the neutron can be deduced from Eqs. (\ref{eq: sigma}) and (\ref{eq: gamma}) by the substitution:
\beq
e_d\lrar e_u~;~~~~~~~\mu_p,~\mu_p^a\rar~\mu_n~; ~~~~~~~A_p,~B_p\rar A_n,~B_n~.
\eeq
Multiplying Eqs. (\ref{eq: sigma}) and (\ref{eq: gamma})  by $e_d$ and each corresponding neutron sum rule by $e_u$ and then subtracting the proton and the corresponding neutron sum rules, IS deduce:
\bea
 \mu_pe_d-\mu_n e_u+(A_pe_d-A_n e_u)\tau^{-1}=\nnb\\
 {4a^2\over 3\tilde\lambda_N^2}e^{M^2_N\tau} (e_u^2-e_d^2)~,\nnb\\
\mu^a_pe_u-\mu_n e_d+(B_pe_u-B_n e_d)\tau^{-1}=\nnb\\
{4aM_N\tau^{-1}\over \tilde\lambda_N^2}e^{M^2_N\tau} (e_u^2-e_d^2)~.
\label{eq:mupn}
\eea
In order to eliminate the single pole contribution, IS apply, to both sides of Eq. (\ref{eq:mupn}),
the operator:
\beq
 {\cal L}_{01}\equiv \ga 1+\tau{\partial\over \partial{\tau}}\dr ~.
\eeq
Using the LSR expression of the proton coupling $\lambda_N$ from the $M_N$-component $F_2$ of the
two-point correlator  \cite{ioffe,dosch,dosch1}:
\beq
\tilde\lambda^2_N\simeq {2a\tau^{-2}\over M_N}e^{M^2_N\tau}~,
\eeq
IS deduces for $\tau^{-1}\approx M^2_N$:
\bea
\mu_p&\approx& {8\over 3}\ga 1+{1\over 6}{a\over M^3_N}\dr\approx 2.96~,\nnb\\
\mu_n&\approx& -{4\over 3}\ga 1+{2\over 3}{a\over M^3_N}\dr\approx -1.93~,
\eea
 in remarkable agreement with the experimental values:
\beq
\mu_p^{exp}=2.79~,~~~~~~~~~\mu_n^{exp}=-1.91~,
\eeq
despite the crude LO approximation used for getting these predictions. Including the OPE and anomalous dimension corrections, IS deduce, from Eq. (\ref{eq:mupn}), the  predictions
\footnote{Some  relations between the neutron anomalous magnetic and its electric dipole moments has been also derived using light-front QCD approach \cite{GARDNER}, which will be interesting to check from some other methods.}:
\beq
\mu_p^{exp}=3.0~,~~~~~~~~~\mu_n^{exp}=-2.0~. 
\eeq
and the correlated value:
\beq
\chi\simeq -8~{\rm GeV}^{-2}~,
\eeq
for $\xi\approx \kappa\approx 0$ . However, by examining the LSR used by the authors, we notice that these sum rules do not satisfy $\tau$-stability criteria such that (a priori) there is no good argument for
extracting an optimal result.\\
In order to  check the previous result, we solve the  two equations:
\beq
\chi={d\over d\tau} {\cal L}^\sigma_{p}~~~~~~{\rm and}~~~~~~~\chi={d\over d\tau} {\cal L}^{\gamma}_{p}~,
\eeq
for  each given value of $t_c$.  The functions ${\cal L}^{\sigma,\gamma}_{p}$ have been defined in Eqs. (\ref{eq: sigma}) and (\ref{eq: gamma}). We use the values of $\kappa$ and $\xi$ given in Eq. (\ref{eq:xikappa}) but they do not affect much the conclusions like in the case of IS. The analysis is shown in Fig. \ref{fig:chioffe}, where a common solution is reached at $\tau=0.4$ GeV$^{-2}$, though the curves do not exhibit  $\tau$-stability region. 
Again like in the proton mass sum rule, the $t_c$-stability is reached around $t_c=3$ GeV$^2$ \cite{dosch1}. Taking as a conservative estimate the range of values $t_c=1.6\sim 3.$ GeV$^2$, where the lowest value corresponds to the beginning of $\tau$-stability for the determination of the proton mass, we deduce the optimal estimate:
\beq
\chi\simeq -(8.5\pm 1.0)~{\rm GeV}^{-2}~,
\label{eq:chisr}
\eeq
in good agreement with the IS previous value \cite{IOFFE2} and the one in \cite{VAIN} using the quark  triangle anomaly. At $\tau=0.4$ GeV$^{-2}$ where a common solution has been obtained, one expects a good convergence of the OPE and smaller effects of radiative corrections. We have used the choice $b=-1$ which we expect to give a reliable result like in the previous cases of the proton mass and $D_N$ discussed in the next section where the results are almost unchanged (within the errors)  from $b=-1$ to the optimal value -1/5 obtained in the case of the proton mass \cite{dosch,dosch1}.
 \subsection*{\b Test of  the  LSR results of Ref. \cite{ritz} for NEDM}
 From the previous QCD and phenomenological expressions of the two-point correlators, one can deduce the Borel/Laplace sum rule (LSR):
 \bea
{\cal L}(\tau)&\equiv&  r(\tau)\equiv{1\over 2 \theta m^*}\ga D_N+{A\tau^{-1}\over \lambda_N^2 M_N}\dr\nnb\\
&=& -{\tau^{-2}\over 32\pi^2} \la \bar\psi\psi\ra e^{M^2_N\tau} \Bigg{\{} \chi C_0(1-\rho_0)+\Bigg{[} C_{2a}\times \nnb\\
&&
\Big{[} -\ln {(\tau\nu_\chi^2)}+\gamma_E-1\Big{]}+C_{2b}\Bigg{]}\tau \Bigg{\}}~,
 \label{eq:ritzsumrule}
  \label{eq:nuritza}
 \eea
 where $\tau^{-1}\equiv M^2$ is the LSR variable.
 Ref. \cite{ritz} uses either the value \cite{ioffe,dosch,dosch1}:
 \beq
 \lambda_N^2\simeq {1\over (2\pi)^4}\ga 1.05\pm 0.1\dr {\rm GeV}^{6}~,
 \label{eq:lambdaIoffe}
 \eeq
 or its LSR expression  \cite{ioffe,dosch,dosch1} from the $\hat q\equiv \gamma_\mu q^\mu$ part of the correlator:
 \bea
 (2\pi)^4\lambda^2_Ne^{-\tau M_N^2} ={5+2b+5b^2\over 64}\tau^{-3}\times \nnb\\\Big{[} 
(1-\rho_2) +\pi\la\alpha_s G^2\ra(1-\rho_0)\tau^{2}\Big{]}~.
 \label{eq:lambda1}
 \eea
 However, due to its high-dependence on $\tau$, this sum rule is much affected by the form of the continuum such that we shall not consider it. 
Instead, we shall consider either the value in Eq. (\ref{eq:lambdaIoffe}), or the expression of the residue from the $M_N$ part of the correlator:
 \bea
 (2\pi)^4\lambda^2_NM_Ne^{-\tau M_N^2} =-{\pi^2\over 4}\la\bar\psi\psi\ra\tau^{-2} \Big{[} (7-2b-5b^2)\nnb\\
 (1-\rho_1)
 -3M^2_0\tau(1-b^2) (1-\rho_0)\Big{]}~,
  \label{eq:lambda2}
 \eea 
 which has a lesser dependence in $\tau$.
 
\begin{center}
\begin{figure}[here]
\includegraphics[width=.22\textwidth]{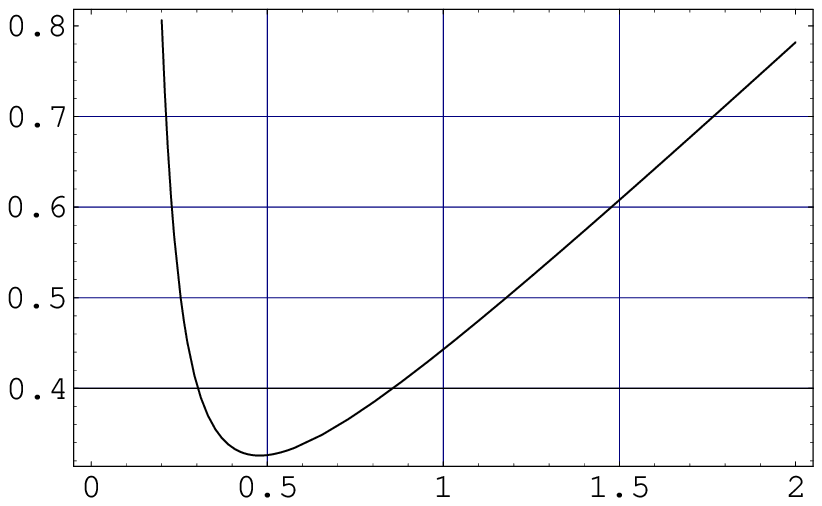}
\includegraphics[width=.22\textwidth]{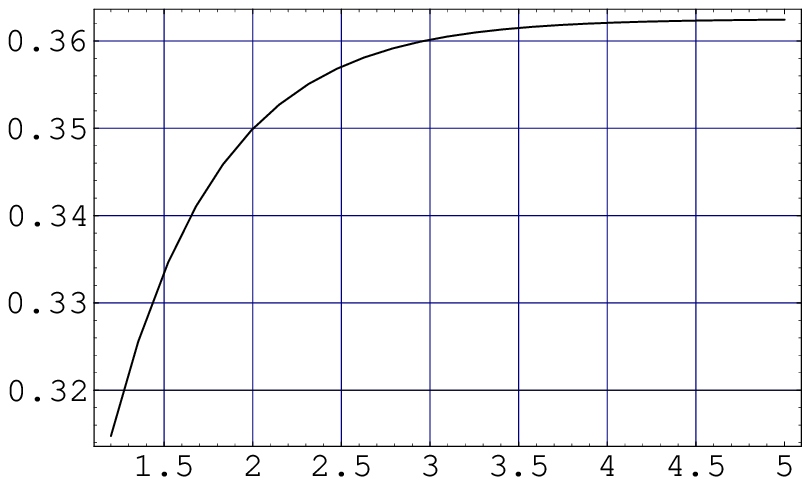}
\caption{Analysis of $-r/\chi$ using LSR for b=1: a) $M^2$-dependence for $t_c=1.6$ GeV$^2$; b) $t_c$-dependence for $M^2=.5$ GeV$^2$.  }
\label{fig:ritz}
\end{figure}
\end{center}

 We show the results in Fig~\ref{fig:ritz} for the previous value of $\lambda_N^2$  and using, as in Ref. \cite{ritz},  $\la  \psi\psi\ra = -[0.225~{\rm GeV}]^3$  for a better comparison.
 
{\bf --}  For the choice $b=1$ used in \cite{ritz}, one can see from Fig~\ref{fig:ritz}a that the optimal value is obtained at $M^2\equiv\tau^{-1} \simeq 0.5$ GeV$^2$, which is relatively low for justifying the convergence of both the OPE and the PT series in $\alpha_s$. Fig~\ref{fig:ritz}b shows, like in the case of the analysis of the proton mass,  that the $t_c$ stability is reached at high-value of 3 GeV$^2$ but the estimate does not move much from the optimal value $t_c=1.6$ GeV$^2$ obtained in the proton mass sum rule \cite{ioffe,dosch,dosch1}. In this case, one can deduce: 
 \beq
 r\vert_{b=1}= -\chi (0.34\sim 0.36)~{\rm GeV}^{4}~,
 \label{eq:nuritz}
 \eeq
 which reproduces the result of \cite{ritz}. Assuming, like in Ref. \cite{ritz}, that the single pole contribution can be neglected
 (which we shall test in the next section), and using the value of $\chi=-5.7$ GeV$^{-2}$ used in \cite{ritz},  one can deduce from Eqs. (\ref{eq:nuritza}) and (\ref{eq:nuritz}) :
\beq
D_N\vert_{b=1}\approx 9 \times 10^{-3}\theta ~{\rm GeV}^{-1}~.
\eeq
 Though (almost) trivial, the previous test is necessary for calibrating our
 sum rule and for checking our inputs in the next analysis.
\subsection*{\b New estimate of $A$ and choice of the nucleon currents}
\begin{center}
\begin{figure*}[hbt]
\includegraphics[width=.33\textwidth]{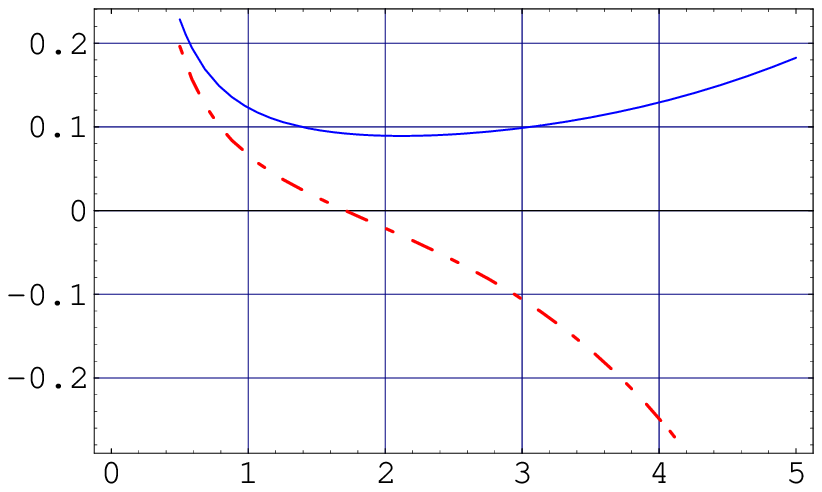}
\includegraphics[width=.33\textwidth]{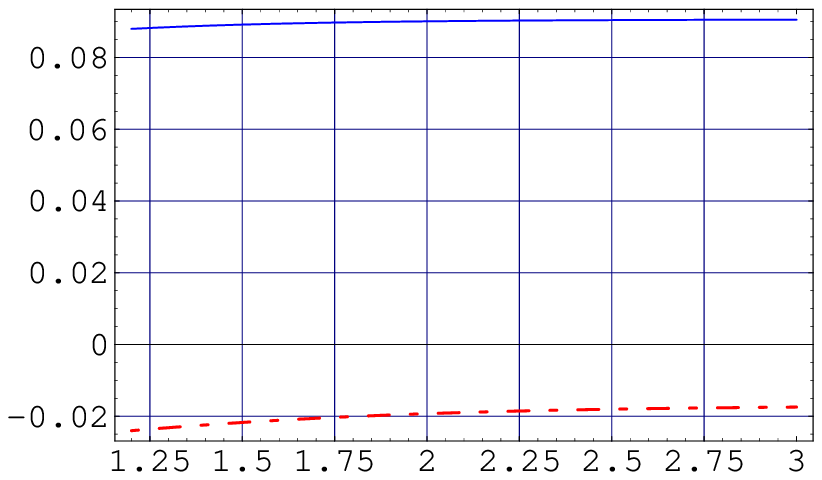}
\includegraphics[width=.33\textwidth]{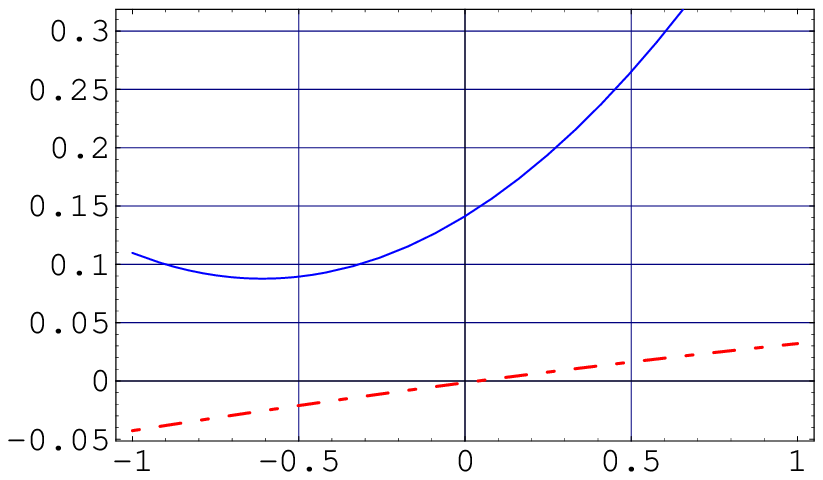}

\caption{ LSR analysis of $-r/\chi$ (red : dash-dotted curve) and $-r_A/\chi$ (blue : continuous curve) : a) $\tau\equiv 1/M^2$-dependence  for $b=-0.5, ~t_c=1.6$ GeV$^2$; b) $t_c$ dependence for  $b=-0.5,~\tau=2$ GeV$^{-2}$; c) $b$-dependence for $t_c=1.6$ GeV$^2$, ~$\tau=2$ GeV$^{-2}$}
\label{fig:Aritz}
\end{figure*}
\end{center}

 We shall reconsider the previous analysis by abandoning the choice $b=1$ for 
 the nucleon current and by giving a new estimate of :
\beq
 r_A(\tau)\equiv {1\over 2 \theta m^*}\ga {A\tau^{-1}\over \lambda_N^2 M_N}\dr~.
 \eeq 
This analysis is summarized in Fig.~\ref{fig:Aritz} where we have used the value of $\lambda_N$ in  
Eq.~(\ref{eq:lambdaIoffe}) and the running condensate value \cite{SNQ,SNB,SNDOSCH}:
\beq
\la 0\vert \psi\psi\vert 0\ra (M)\simeq  -[0.266~{\rm GeV}]^3\ln^{4/9}{\ga M/ \Lambda\dr}~,
\eeq 
with $\Lambda\simeq 350$ MeV for 3-flavours. 

{\bf --} One can notice that the result is optimal in $b$ for:
 \beq
 b\simeq -0.5~,
 \eeq 
 and more conservatively in the range:
 \beq
 -1\leq b \leq 0~,
 \eeq
 which does not favour the choice $b=1$ used in \cite{ritz}. The previous range includes the conventional choices:  -1 in \cite{ioffe}, -1/5 in \cite{dosch,dosch1} and the non-relativistic limit $b=0$ used in lattice calculations \cite{LEIN}.

{\bf --} The 2nd (important) assumption used in Ref. \cite{ritz} is the neglect of the contribution of the single pole controlled by the parameter $A$. By inspecting the LSR in Eq. (\ref{eq:ritzsumrule}), one can isolate $A$ by working with the new LSR:
\beq
{\cal L}_1(\tau)\equiv {d\over d\tau} {\cal L}~.
\eeq
One can see in Fig. \ref{fig:Aritz}c that for all  ranges of $b$, $r_A$ is much smaller than $r$ justifying the assumption 
of \cite{ritz}. At the optimal range of $b$ values given previously, one can deduce :
\beq
{r/ \chi}=- 0.09~,~~~~~~~~~~~{r_A/ \chi}=+0.02~.
\eeq
Using the quark mass values in \cite{SNQ,SNB} and the previous value of $\chi$ in Eq. (\ref{eq:chisr}), one gets:
\beq
D_N\simeq -0.11\chi\theta m^*\simeq (2.24\pm 0.12) \times 10^{-3}\theta ~{\rm GeV}^{-1}~.
\eeq

\subsection*{\b Direct extraction of $D_N$ from a new LSR}
\begin{center}
\begin{figure*}[hbt]
\includegraphics[width=.33\textwidth]{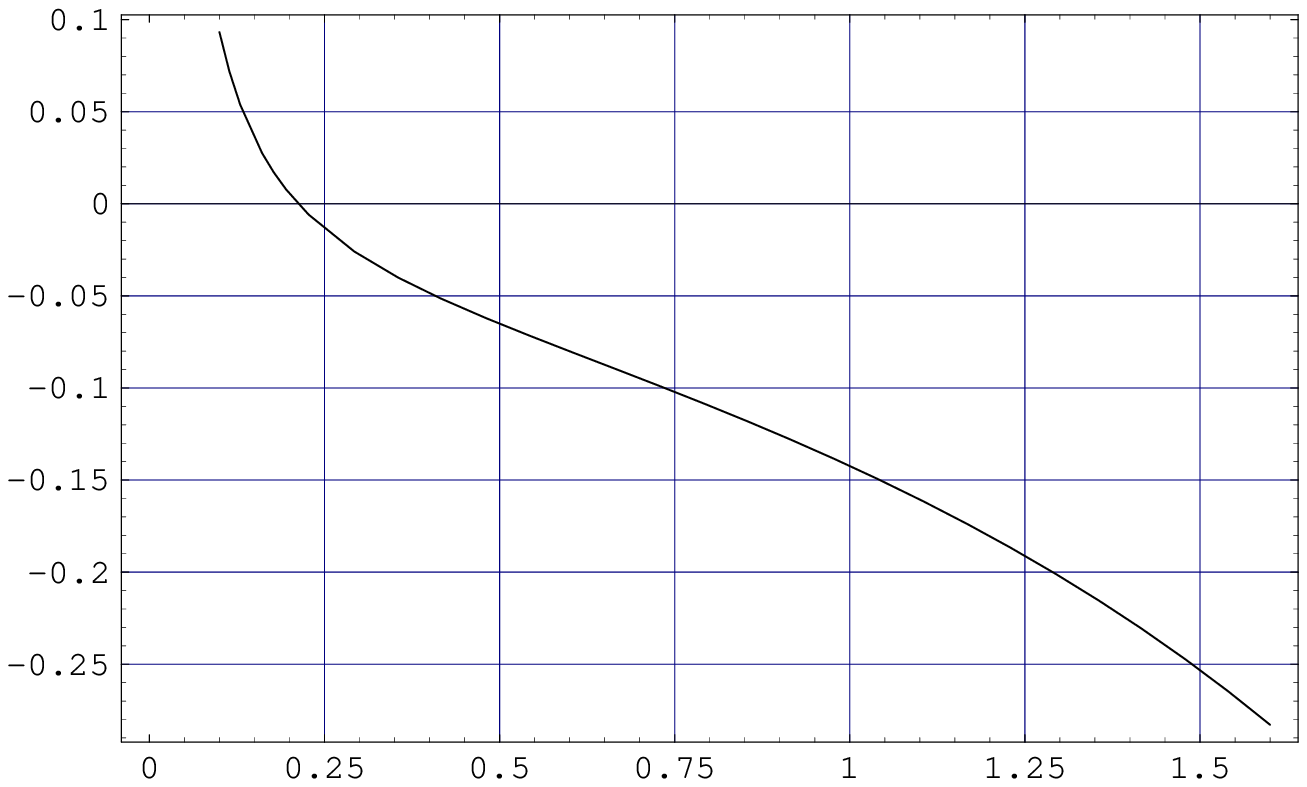}
\includegraphics[width=.33\textwidth]{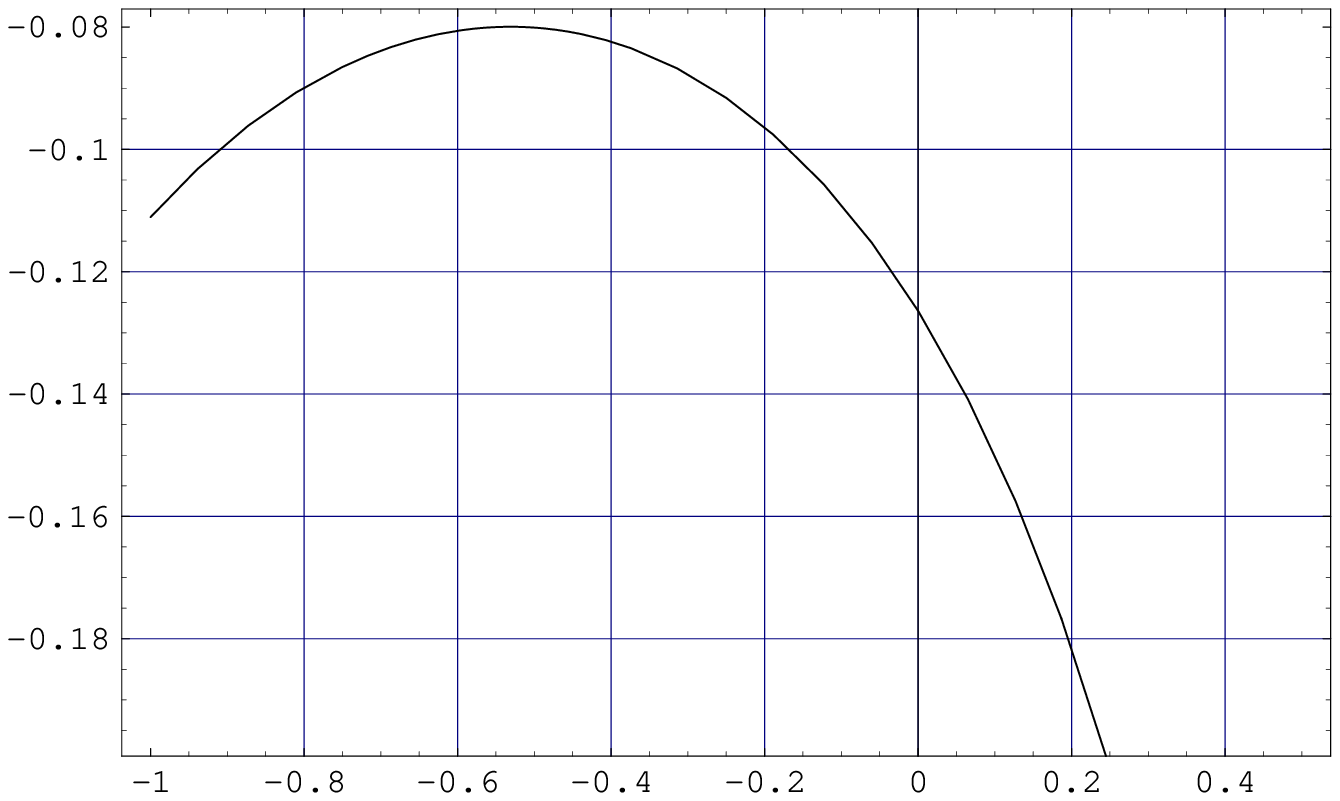}
\includegraphics[width=.33\textwidth]{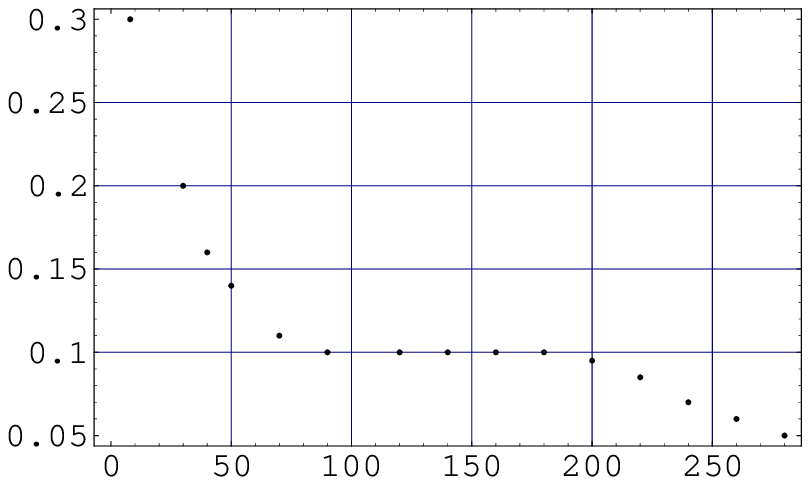}
\caption{Analysis of $D_N$ using LSR a) $\tau\equiv 1/M^2$ in GeV$^{-2}$-dependence for $b=-0.5, ~t_c=1.6$ GeV$^2$; b)~$b$-dependence for $\tau=0.75$ GeV$^{-2}$, $~t_c=1.6$ GeV$^2$; c)~$\nu_\chi$ -dependence of the optimal value in units of MeV.}
\label{fig:dnsr}
\end{figure*}
\end{center}

{\bf --} By inspecting the LSR in Eq. (\ref{eq:ritzsumrule}), one can also isolate  $D_N$ by working with the LSR:
\beq
{\cal L}_2(\tau)\equiv {d\over d\tau} \tau{\cal L}~.
\eeq
We show the results of the analysis in Fig.~\ref{fig:dnsr}.
One can notice that the sum rule stabilizes at $\tau\approx 0.75$ GeV$^{-2}$, which is smaller than in the 
previous analysis, showing a better convergence of the OPE.

{\bf --} We also study the dependence of the result on the value of the IR scale $\nu_\chi$. The optimal value corresponds to:
\beq
\nu_\chi\simeq (80\sim 200)~{\rm MeV}~,
\label{eq:nuchi}
\eeq
which has the size of a typical IR chiral scale (pion or constituent quark mass).

Therefore, we deduce the optimal value:
\bea
D_N&\simeq& -(0.10 \pm 0.03\pm 0.03)\chi\theta m^*\nnb\\
&\simeq& (2.06 \pm 0.08)\times 10^{-3}\theta ~{\rm GeV}^{-1}~,
\eea
where the first (resp. second) error comes from the localization of the extremum in $\tau$ (resp.) of the $t_c$ values
which we take to be $t_c=(1.6\pm 0.2)$ GeV$^2$ around the value obtained from the proton mass sum rule \cite{dosch,dosch1}.
\subsection*{\b NEDM results and systematics from QSSR}
The two results from the LSR are in good agreement and lead to the
final estimate:
\beq
D_N\vert_{\rm QSSR}\simeq  (2.15\pm 0.10) \times 10^{-3}\theta ~{\rm GeV}^{-1}~,
\eeq
which agrees with the range spanned by the chiral and ChPT estimate in Eq. (\ref{eq:dnchiral}). 

In order to analyze the systematic errors in the approach,
we estimate using vertex sum rules the well measured coupling  $g_{\pi NN}=13.4$. 
We use the symmetric configuration of the hadronic vertex in \cite{PAVER} from which,
one obtains the LSR:
\bea
g_{\pi NN}(\lambda_N^2M_N)(f_\pi m_\pi^2)\tau^3 e^{-M^2_N\tau}\simeq \nnb\\
-{1\over 16\pi^2}(m_u+m_d)\la\bar \psi\psi\ra
\ga 2+8b+8b^2\dr~,
\eea
Using the expression of $\lambda_N^2$ in Eq. (\ref{eq:lambda1}), one can deduce the LO sum rule:
\beq
g_{\pi NN}\simeq {16\pi^2f_\pi\over M_N}\ga {2+8b+8b^2\over 5+2b+5b^2}\dr ~,
\label{eq:gpiNN}
\eeq
while the one in Eq. (\ref{eq:lambda2}) gives:
\beq
g_{\pi NN}\simeq {(m_u+m_d)\over f_\pi m_\pi^2} \tau^{-1}\ga {2+8b+8b^2\over 7-2b-5b^2}\dr ~.
\eeq
Using a double pole dominance and neglecting the QCD continuum, Ref. \cite{PAVER} fixes from
Eq. (\ref{eq:gpiNN}) the nucleon operator mixing to be $b\equiv 1/t=0.307$ for the 1st sum rule to reproduce the 
experimental value of $g_{\pi NN}$. One can notice that this sum rule is not accurate due to its 
$\tau$-dependence. For $\tau\simeq  1$ GeV$^{-2}$, the previous value of the quark mass evaluated
at 1 GeV is $(m_u+m_d)(1)\simeq$ 10.9 MeV. Including the QCD continuum contribution with $t_c=1.6$ GeV$^2$, the 2nd sum rule gives:
\beq
g_{\pi NN}\simeq 8.9~.
\label{eq: qssrpinn}
\eeq
We consider its deviation by 33\% from the data as the systematic error of QSSR for this estimate
\footnote{Note that a more precise estimate of $g_{\pi NN}$ including the contribution of the two first lowest quark and gluon condensate contributions is claimed in Ref. \cite{RRY}  from the 1st sum rule
using a different configuration of the hadronic vertex.}.
Therefore, we consider as a conservative estimate of  $D_N$ from QSSR:
\bea
D_N\vert_{\rm QSSR}&\simeq&  (2.15\pm 0.71) \times 10^{-3}\theta ~{\rm GeV}^{-1}\nnb\\
&\simeq& (4.24\pm 1.40) \times 10^{-17}\theta ~{\rm cm}~.
\label{eq:dnQSSR}
\eea
\section{Constituent quark results}
For a qualitative
comparison of the results from the chiral and QSSR approaches, we use a simple model where the constituent quark interacts with electromagnetic field. Then, one can write \cite{IOFFE2}:
\bea
\la 0\vert \bar\psi \sigma_{\mu\nu} \psi \vert 0\ra \vert_{F}&\equiv &e_q\chi F_{\mu\nu}\la 0\vert \bar\psi\psi\vert 0\ra\nnb\\
&=& -\int d^4p {Tr} \aga S(p,M_q)\sigma_{\mu\nu}\adr~,
\eea
where $S(p,M_q)$ is the quark propagator in presence of an electromagnetic field:
\bea
S(p,M_q)&=& {i\over (2\pi)^4}\Bigg{[} {1\over \hat p-M_q}-{1\over 2}\la 0\vert \bar\psi \sigma_{\mu\nu} \psi \vert 0\ra \vert_{F}\nnb\\
&&\Big{[}{i\over \hat p-M_q}\gamma_\mu {1\over \hat p-M_q}\gamma_\nu {1\over \hat p-M_q}\nnb\\
&&-{\mu_a^a\over 2M_q}{1\over \hat p-M_q}\sigma_{\mu\nu}{1\over \hat p-M_q}\Big{]}\Bigg{]}~,
\eea
where $M_q$ the quark constituent mass and  $\mu_q^a\simeq 2$ is its anomalous magnetic moment. 
Then, one can derive the relation \cite{IOFFE2}:
\beq
\chi \la 0\vert \psi\psi\vert 0\ra = {3\over 2\pi^2} M_q \ln \ga {\nu\over  M_q} \dr\ga 1+{\mu_q^a\over 2}\dr~.
\eeq
Using this relation into the LSR expression of $D_N$, one can deduce in units of $e$:
\bea
D_N&\approx& 3 M_q \ln{ \ga\nu\over M_q\dr} \tau^{-2} e^{M^2_N\tau}\ga 1 +{\mu_q^a\over 2}\dr
\theta m^*\nnb\\
&\approx& 4.4 \times 10^{-3} \theta ~{\rm GeV}^{-1}\approx  8.7 \times 10^{-17} \theta ~{\rm cm},
\eea
where we have taken $b=0$ in the non-relativistic limit, $\tau^{-1}\approx M^2_N\approx \nu^2$,  and we have used $M_q\approx  (200\sim 300)$ MeV $\approx \nu_\chi$. We assume that this crude non-relativistic approximation is known with an accuracy of about 50\%, which gives the final estimate:
\beq
D_N\vert_{\rm Const~quark}\approx (8.7\pm 4.4) \times 10^{-17} \theta ~{\rm cm}~.
\label{eq:dnconst}
\eeq
This value can be compared with the one from more involved LSR analysis. This approximate formula may indicate that $D_N$ is dominated by the non-analytic Log. contribution like in the case of the chiral estimate of \cite{CVW} rederived in the previous section, but at the quark constituent level.\\

\section{ Final range of the NEDM values }
The previous results from chiral and ChPT approaches in Eqs. (\ref{eq:dnchiral}), from QSSR in Eq. (\ref{eq:dnQSSR}) and from a na\"\i ve quark constituent model in Eq. (\ref{eq:dnconst}) are comparable. However, a more definite comparison with the chiral and ChPT estimate requires a better control of the value of the renormalization scale $\nu$ and an improved estimate of the $CP$ violating $\pi NN$ coupling. Also, search for some other contributions beyond the standard OPE of QSSR like e.g.
the one of the $D=2$ dimension operator discussed \cite{CHET}, may be required.

Combining these previous results with the present experimental upper limit (in units of e) \cite{EXP}:
\beq
D_N\vert_{\rm exp}\leq 6.3\times 10^{-26}  ~{\rm cm}~,
\eeq
one can deduce in units of $10^{-10}$:
\bea
\theta& \leq& (1.6\pm  0.4) ~{\rm [Chiral] }~:~\nu=M_N\nnb\\
& \leq& (1.3\sim  11.7) ~{\rm [ChPT] }~:~M_N/3\leq \nu \leq M_N\nnb\\
&\leq& (6.9\pm 3.5) ~~{\rm [Constituent~quark] }\nnb\\
&\leq& (14.9\pm 4.9) ~~{\rm [QSSR] }~.
\eea
These results indicate that the weakest upper bound comes from QSSR, while the strongest upper bound comes from the chiral estimate evaluated at the scale $\nu=M_N$. Present lattice calculations are at an early stage \cite{LATTICE} and may narrow the previous range of values in the future. 
\section*{Acknowledgement}
It is a pleasure to thank S. Friot, P. Di Vecchia and G. Veneziano for collaboration in deriving some of the results in Section 2 and for multiple email exchanges. Communications with E. de Rafael and V.I. Zakharov are also appreciated. This work has been initiated when the author has visited the CERN Theory Group in autumn 2006, which he wishes to thank for its hospitality.

\end{document}